\documentstyle[aps]{revtex}
\begin{document}
\newcommand{\s}{\varsigma}
\newcommand{\room}{\rule[-0.3cm]{0cm}{0.8cm}}
\newcommand{\hsp}{\hspace*{3mm}}
\newcommand{\vsp}{\vspace*{3mm}}
\newcommand{\nsp}{\vspace*{-3mm}}
\newcommand{\be}{\begin{equation}}
\newcommand{\ee}{\end{equation}}
\newcommand{\bd}{\begin{displaymath}}
\newcommand{\ed}{\end{displaymath}}
\newcommand{\bdm}{\begin{displaymath}}
\newcommand{\edm}{\end{displaymath}}
\newcommand{\bea}{\begin{eqnarray}}
\newcommand{\eea}{\end{eqnarray}}
\newcommand{\sgn}{~{\rm sgn}}
\newcommand{\extr}{~{\rm extr}}
\newcommand{\Equiv}{\Longleftrightarrow}
\newcommand{\pprime}{\prime\prime}
\newcommand{\notexists}{\exists\hspace*{-2mm}/}
\newcommand{\bra}{\langle}
\newcommand{\ket}{\rangle}
\newcommand{\bigbra}{\left\langle\room}
\newcommand{\bigket}{\right\rangle\room}
\newcommand{\bras}{\langle\!\langle}
\newcommand{\kets}{\rangle\!\rangle_{xy}}
\newcommand{\bigbras}{\left\langle\!\!\!\left\langle\room}
\newcommand{\bigkets}{\right\rangle\!\!\!\right\rangle_{\!\!xy}\room}
\newcommand{\order}{{\cal O}}
\newcommand{\minus}{\!-\!}
\newcommand{\plus}{\!+\!}
\newcommand{\erf}{{\rm erf}}
\newcommand{\bk}{\mbox{\boldmath $k$}}
\newcommand{\bm}{\mbox{\boldmath $m$}}
\newcommand{\br}{\mbox{\boldmath $r$}}
\newcommand{\bq}{\mbox{\boldmath $q$}}
\newcommand{\bz}{\mbox{\boldmath $z$}}
\newcommand{\bH}{\mbox{\boldmath $H$}}
\newcommand{\bM}{\mbox{\boldmath $M$}}
\newcommand{\bQ}{\mbox{\boldmath $Q$}}
\newcommand{\bR}{\mbox{\boldmath $R$}}
\newcommand{\bW}{\mbox{\boldmath $W$}}
\newcommand{\hbh}{\hat{\mbox{\boldmath $h$}}}
\newcommand{\hbm}{\hat{\mbox{\boldmath $m$}}}
\newcommand{\hbr}{\hat{\mbox{\boldmath $r$}}}
\newcommand{\hbq}{\hat{\mbox{\boldmath $q$}}}
\newcommand{\hbD}{\hat{\mbox{\boldmath $D$}}}
\newcommand{\hbQ}{\hat{\mbox{\boldmath $Q$}}}
\newcommand{\hbR}{\hat{\mbox{\boldmath $R$}}}
\newcommand{\hbW}{\hat{\mbox{\boldmath $W$}}}
\newcommand{\bsigma}{\mbox{\boldmath $\sigma$}}
\newcommand{\bomega}{\mbox{\boldmath $\Omega$}}
\newcommand{\bphi}{\mbox{\boldmath $\Phi$}}
\newcommand{\bpsi}{\mbox{\boldmath $\psi$}}
\newcommand{\bdelta}{\mbox{\boldmath $\Delta$}}
\newcommand{\btheta}{\mbox{\boldmath $\theta$}}
\newcommand{\bxi}{\mbox{\boldmath $\xi$}}
\newcommand{\bmu}{\mbox{\boldmath $\mu$}}
\newcommand{\brho}{\mbox{\boldmath $\rho$}}
\newcommand{\bEta}{\mbox{\boldmath $\eta$}}
\newcommand{\req}{r_{\rm eq}}
\newcommand{\unity}{{\bf 1}\hspace{-1mm}{\bf I}}

\title{\bf Dynamical Replica Theory for Disordered Spin Systems}
\author{A.C.C. Coolen\footnote{present address: Dept. of Mathematics,
King's College London, Strand, London WC2R 2LS, U.K.}
\and S.N. Laughton \and D. Sherrington}
\maketitle

\begin{center}
Department of Physics - Theoretical Physics\\
University of Oxford\\
1 Keble Road, Oxford OX1 3NP, U.K.
\end{center}

\begin{center} PACS: 75.10.Nr, 05.20.-y \end{center}\vsp

\begin{abstract}\noindent
We present a new method to solve the dynamics of
disordered spin systems on finite time-scales.
It involves a
closed driven diffusion
equation for the joint spin-field distribution,
with time-dependent
coefficients described by a dynamical replica theory
which, in the case of detailed balance,
incorporates equilibrium replica theory as a stationary state.
The theory is exact in various limits.
We apply our theory to both the symmetric- and
the non-symmetric Sherrington-Kirkpatrick spin-glass, and show that it
describes the (numerical) experiments very well.
\end{abstract}
\vsp\vsp

Recently it has become clear \cite{cuku1} that even mean-field
models exhibit the
ageing phenomena, familiar from experiments on real spin-glass
\cite{vincentetal}, that where hitherto assumed to be typical for
short-range systems. This has led to a renewed interest in
dynamical studies of mean-field spin-glass models and to valuable new
insights into spin-glass dynamics away from equilibrium,
see e.g. \cite{cuku2}.
In this letter we present a novel
approach to analysing the dynamics of
spin-glass models on finite time-scales, leading to a {\rm
dynamical} replica theory, which, in the case of detailed balance,
incorporates equilibrium replica theory as a stationary state
(including replica symmetry breaking, if it occurs).  The formalism is built on
a closure procedure
which which we obtain a closed diffusion equation for the joint
spin-field distribution.
It constitutes the fixed-point of a series of
previous dynamical studies \cite{CS,sk1,CF}.
Our theory is proven to be exact for short times and in
equilibrium. For intermediate times we can prove that it
is exact if in the thermodynamic limit the joint
spin-field distribution indeed obeys a closed dynamic equation.
Here we discuss only the underlying physical ideas and the
results of applying our theory to both the symmetric-
\cite{SK} and the non-symmetric \cite{crisantisompolinsky}
Sherrington-Kirkpatrick spin-glass. Full mathematical details will be
published elsewhere \cite{LCS}.  We believe the
agreement between theory and (simulation) experiment to be quite
convincing.
\vsp

The generalised (asymmetric) version of the Sherrington-Kirkpatrick
(SK) model \cite{SK}, introduced in \cite{crisantisompolinsky}, consists of $N$
Ising spins
$\sigma_{i}\in\{-1,1\}$ with infinite-range interactions $J_{ij}$:
\be
J_{ij}=\frac{J_0}{N}+\frac{J}{\sqrt{N}}\left[\cos(\frac{1}{2}\omega)x_{ij}\plus\sin(\frac{1}{2}\omega)y_{ij}\right]
{}~~~~~~
x_{ij}=x_{ji},~~y_{ij}=-y_{ji}
\label{eq:interactions}
\ee
For $i<j$ each of the random quantities $x_{ij}$ and $y_{ij}$,
representing
quenched disorder, are  drawn independently from a Gaussian distribution with
zero mean and unit variance.
The evolution in time of the microscopic probability distribution
$p_{t}(\bsigma)$ is given by
the master equation
\be
\frac{d}{dt}p_{t}(\bsigma)=\sum_{k=1}^{N}\left[p_{t}(F_k\bsigma)w_{k}(F_k\bsigma)-p_{t}(\bsigma)w_{k}(\bsigma)\right]
\label{eq:master}
\ee
in which $F_k$ is a spin-flip operator
$F_{k}\Phi(\bsigma)\equiv\Phi(\sigma_1,\ldots,-\sigma_k,\ldots,\sigma_N)$ and
the transition rates $w_k(\bsigma)$ and the local alignment fields
$h_i(\bsigma)$ are
\be
w_{k}(\vec{s})=\frac{1}{2}\left[1-\sigma_k\tanh[\beta
h_k(\bsigma)]\right]~~~~~~~~
h_{i}(\bsigma)=\sum_{j\neq i}J_{ij}\sigma_{j}+\theta
\label{eq:ratesandfields}
\ee
where $\beta=1/T$ is the inverse temperature. The mixing angle
$\omega\in[0,\pi]$
controls the degree of symmetry of the interactions
(\ref{eq:interactions}). For $\omega=0$ we recover the original SK
spin-glass model \cite{SK}. Now the interactions are symmetric,
the dynamics obeys detailed balance and (\ref{eq:master}) reduces to a Glauber
dynamics, leading asymptotically to the Boltzmann equilibrium
distribution $p_\infty(\bsigma)\sim \exp[\minus\beta H(\bsigma)]$
with the conventional Hamiltonian
\be
H(\bsigma)= -\sum_{i< j}\sigma_i J_{ij}\sigma_j-\theta\sum_i\sigma_i
\label{eq:hamiltonian}
\ee
For $\omega>0$, however, detailed balance is violated and equilibrium
statistical mechanics no longer applies.
For $\omega=\pi$ the interaction matrix is fully
anti-symmetric.
\vsp

For any given set of $\ell$ macroscopic
observables $\bomega (\bsigma)= ( \Omega_1(\bsigma), \dots,
\Omega_\ell(\bsigma))$ we can derive a macroscopic stochastic
equation in the form of a Kramers-Moyal expansion.
For deterministically evolving observables (in the thermodynamic
limit) on finite time-scales only the first (Liouville) term in this expansion
will survive,
leading to the deterministic flow equation
\be
\frac{d}{dt} \bomega_t = \lim_{N\rightarrow\infty}
\frac{\sum_{\bsigma} p_t (\bsigma)\delta\left[\bomega \minus \bomega
(\bsigma)\right]
\sum_i w_i (\bsigma)\left[
\bomega(F_i\bsigma)\minus \bomega(\bsigma) \right]}
{\sum_{\bsigma} p_t (\bsigma) \delta\left[\bomega \minus \bomega
(\bsigma)\right]}
\label{eq:omegaflow}
\ee
There are two {\em natural} ways for (\ref{eq:omegaflow}) to
close.
Firstly, by the argument of the subshell average in
(\ref{eq:omegaflow}) depending on
$\bsigma$ only through $\bomega(\bsigma)$ (now $p_t(\bsigma)$ will drop out),
and secondly by the microscopic dynamics
(\ref{eq:master}) allowing for equipartitioning solutions (where
$p_t(\bsigma)$ depends on $\bsigma$ only through $\bomega(\bsigma)$).
In both cases the correct flow equation are obtained
upon simply eliminating $p_t(\bsigma)$ from (\ref{eq:omegaflow}).
We now make two assumptions:
$(i)$ the observables $\bomega(\bsigma)$ are self-averaging with respect to
the microscopic realisation of the disorder $\{x_{ij},y_{ij}\}$, at any time,
and $(ii)$ in evaluating the sub-shell average we assume equipartitioning
of probability within the $\bomega$-subshell of the ensemble.
As a result the macroscopic
equation (\ref{eq:omegaflow}) is replaced
by a closed one, from which the unpleasant fraction is subsequently removed
with
a replica identity (see e.g. \cite{KS}):
\be
\frac{d}{dt} \bomega_t = \lim_{N\rightarrow\infty}\lim_{n\rightarrow0}
\sum_{\bsigma^1}\cdots\sum_{\bsigma^n}
\sum_i
\left\langle
w_i (\bsigma^1)\left[
\bomega(F_i\bsigma^1)\minus \bomega(\bsigma^1) \right]
\prod_{\alpha=1}^n \delta\left[\bomega \minus \bomega
(\bsigma^\alpha)\right]
\right\rangle_{\{x,y\}}
\label{eq:closedomegaflow}
\ee
For observables truly governed by a closed equation our closure
procedure reduces to the natural one (in the sense discussed above),
so we know by construction: if a closed self-averaging equation for
$\bomega_t(\bsigma)$ exists, it must indeed be (\ref{eq:closedomegaflow}).
For the set of observables $\bomega(\bsigma)$ we now choose the (infinite
dimensional) joint
spin-field distribution:
\be
D(\s,h;\bsigma)
= \frac{1}{N} \sum_i \delta_{\s,\sigma_i}
\delta\left[h \minus h_i (\bsigma)\right]
\label{eq:distribution}
\ee
with the local fields (\ref{eq:ratesandfields}). Since both the magnetisation
$m=\frac{1}{N}\sum_i\sigma_i$ and the energy
(\ref{eq:hamiltonian}) can be written as integrals over
$D(\s,h;\bsigma)$, the theory will
 automatically be exact $(i)$ in the limit $J\rightarrow0$, $(ii)$
for
$t\rightarrow0$ (upon choosing appropriate initial conditions)
and $(iii)$ in the limit $t\rightarrow\infty$ (for systems evolving towards
equilibrium).
To circumvent technical subtleties we assume that the distribution
(\ref{eq:distribution}) is sufficiently well behaved:
we evaluate $D_t(\s,h)$ in a number $\ell$ of field arguments
and take the limit
$\ell\rightarrow\infty$ {\em after} the limit
$N\rightarrow\infty$. A closed diffusion equation for $D_t(\s,h)$
was also derived in
\cite{horner}. Although similar in spirit to ours, the latter study employed a
different closure procedure, lacking the properties of the present one
of built-in exactness in various limits.
\vsp

We can now run the familiar machinery of replica theory and evaluate
(\ref{eq:closedomegaflow}) for the choice (\ref{eq:distribution}).
The distribution $D_t(\s,h)$ can be shown to indeed evolve deterministically.
Details of this exercise, as usual involving a saddle-point problem, can be
found in \cite{LCS}. The
result is  the diffusion equation
\bd
\frac{\partial}{\partial t} D_t(\s,h) =
\frac{1}{2}\left[1\plus\s\tanh(\beta h)\right]D_t(\minus\s,h)
-\frac{1}{2}\left[1\minus\s\tanh(\beta h)\right]D_t(\s,h)
\ed
\nsp
\be
+\frac{\partial}{\partial h} \left\{D_t(\s,h)
\left[h\minus\theta\minus J_0\bra\tanh(\beta
H)\ket_{D_t}\right]
+ {\cal A}[\s,h;D_t]
+J^2\left[1\minus\bra\sigma\tanh(\beta
H)\ket_{D_t}\right]\frac{\partial}{\partial h}D_t(\s,h)\right\}
\label{eq:finaldiffusion}
\ee
with the short-hand $\bra
f(\sigma,H)\ket_D=\sum_\sigma\int\!dH~D(\sigma,H)f(\sigma,H)$.
We find all interesting physics to be concentrated in a single driving term
${\cal A}$:
\bd
{\cal A}[\s,h;D]=
\minus\lim_{n\rightarrow 0}\sum_{\alpha\beta}(q^{-1})_{\alpha\beta}
\left\{
\bra \tanh(\beta H_1)\sigma_\alpha \ket_M ~
\bra \delta[h\minus H_1]\delta_{\s,\sigma_1}
[H_\beta\minus \theta\minus J_0 m\plus iJ^2\cos\omega(\bR^{\dag}\bsigma)_\beta]
\ket_M
\right.
\ed
\nsp
\be
\left.
+~\cos\omega ~
\bra \delta[h\minus H_1]\delta_{\s,\sigma_1}\sigma_\alpha\ket_M~
\bra\tanh(\beta H_1)
[H_\beta\minus \theta\minus J_0 m\plus iJ^2\cos\omega(\bR^{\dag}\bsigma)_\beta]
\ket_M~
\right\}
\label{eq:rsbintensivepart}
\ee
This involves an effective measure $M$ in replica space:
\bd
\bra f[\bH,\bsigma] \ket_M=
\frac{\int\!d\bH\sum_{\bsigma}
M[\bH,\bsigma]f[\bH,\bsigma]}
{\int\!d\bH\sum_{\bsigma} M[\bH,\bsigma]}
\ed
\bd
M[\bH,\bsigma]=\exp\left\{
- i\hbm\cdot\bsigma - \frac{1}{2}J^2\bsigma\cdot\bQ\bsigma
- i\sum_{\alpha}\hat{D}_{\alpha}(\sigma_\alpha,H_\alpha)
{}~~~~~~~~~~~~~~~~~~~~~~~~~~~~~~~~~~~~~~
\right.
\ed
\nsp
\be
\left.
{}~~~~~~~~~~ - \frac{1}{2J^2}
[\bH\minus \btheta\minus J_0 \bm\plus
iJ^2\cos\omega\bR^{\dag}\bsigma]\cdot\bq^{-1}
[\bH\minus \btheta\minus J_0 \bm\plus iJ^2\cos\omega\bR^{\dag}\bsigma]
\right\}
\label{eq:effectivemeasure}
\ee
with $\bH=(H_1,\ldots,H_n)$ and $\bsigma=(\sigma_1,\ldots,\sigma_n)$.
The $n\times n$ matrices $\{\bq,\bQ,\bR\}$, the $n$-vectors
$\{\hbm,\bm\}$ and the functions $\hat{D}_\alpha(\sigma,H)$ are
obtained by extremisation of the surface $\Psi$:
\bd
\Psi=
i\sum_{\alpha\sigma}\int\!dH~D(\sigma,H)\hat{D}_\alpha(\sigma,H)
+ i\sum_\alpha m_\alpha \hat{m}_\alpha
+\frac{1}{2}J^2\sum_{\alpha\beta} \left[
q_{\alpha\beta}Q_{\alpha\beta}\plus \cos\omega R_{\alpha\beta}R_{\beta\alpha}
\right]
\ed
\nsp
\be
-\frac{1}{2}\log{\rm det}~\bq
+\log \int\!d\bH\sum_{\bsigma}M[\bH,\bsigma]
\label{eq:rsbpsi}
\ee
For the detailed balance case $\omega=0$ (the SK \cite{SK} model), the
equilibrium state calculated within equilibrium statistical
mechanics \cite{parisi},
is found to define a stationary state of
(\ref{eq:finaldiffusion}).
Here we restrict ourselves to replica-symmetric (RS)
saddle-points of $\Psi$, an analysis of (\ref{eq:rsbpsi}) involving
broken replica  symmetry (RSB) \'{a} la Parisi \cite{parisi} is the
subject of a future study \cite{sk3}.
It is a straightforward bookkeeping exercise to
work out the RS saddle-point equations and the corresponding expression ${\cal
A}_{\rm RS}$ for the driving term
(\ref{eq:rsbintensivepart}), which, upon insertion into
(\ref{eq:finaldiffusion}), controls the evolution of the joint
spin-field distribution in RS approximation.
Evaluating ${\cal A}_{\rm RS}$ requires
solving the RS saddle-point equations at each instance of time.
In the usual manner one can also calculate the AT instability \cite{AT}
with respect to replicon fluctuations; here involving variation of
three order parameter matrices, as opposed to one.
\vsp

\pagebreak
\begin{figure}[t]
\vspace*{80mm}
\hbox to
\hsize{\hspace*{-0cm}\includegraphics{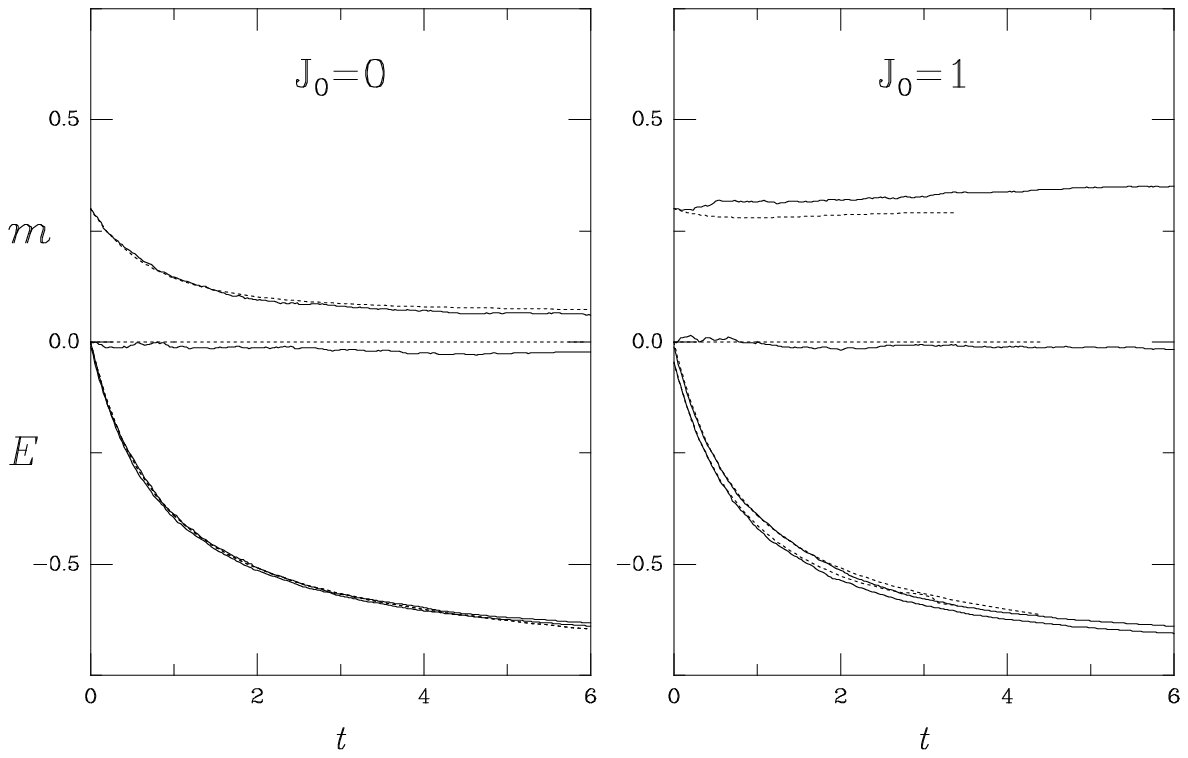}\hspace*{0cm}}
\vspace*{-5mm}
\caption{Magnetisation $m$ and energy
per spin $E$ in the SK model at $T=0$, for $J_0=0$ (left) and $J_0=1$ (right) .
Solid lines:
numerical simulations with $N=8000$; dotted lines: result of solving
the RS diffusion equation.}
\label{fig:flowT0}
\vspace*{105mm}
\hbox to
\hsize{\hspace*{-0cm}\includegraphics{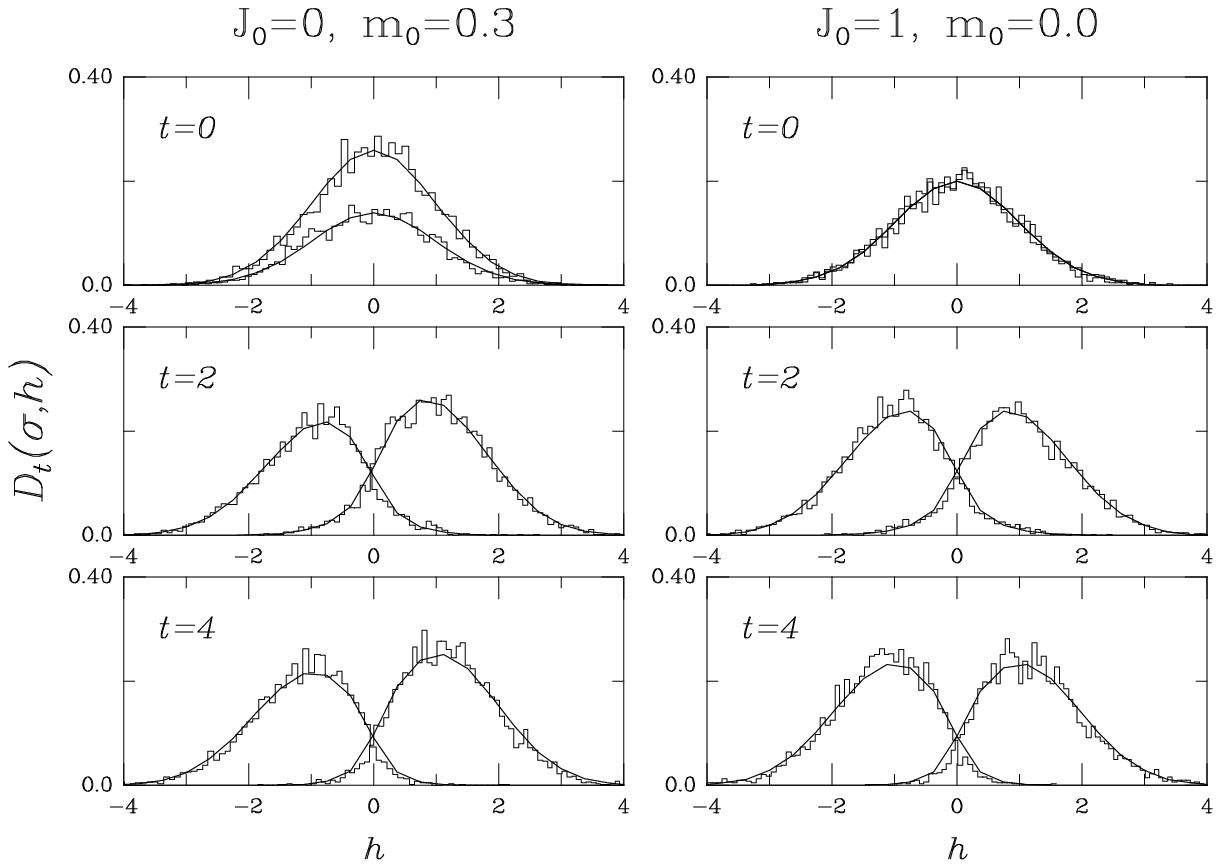}\hspace*{0cm}}
\vspace*{-5mm}
\caption{Field distribitions $D_t(\sigma,h)$ in the SK model at $T=0$, for
$J_0=0$ (left) and $J_0=1$ (right) . Histograms:
numerical simulations with $N=8000$; lines: result of solving
the RS diffusion equation.}
\label{fig:distT0}
\end{figure}
\clearpage
\begin{figure}[t]
\vspace*{80mm}
\hbox to
\hsize{\hspace*{-0cm}\includegraphics{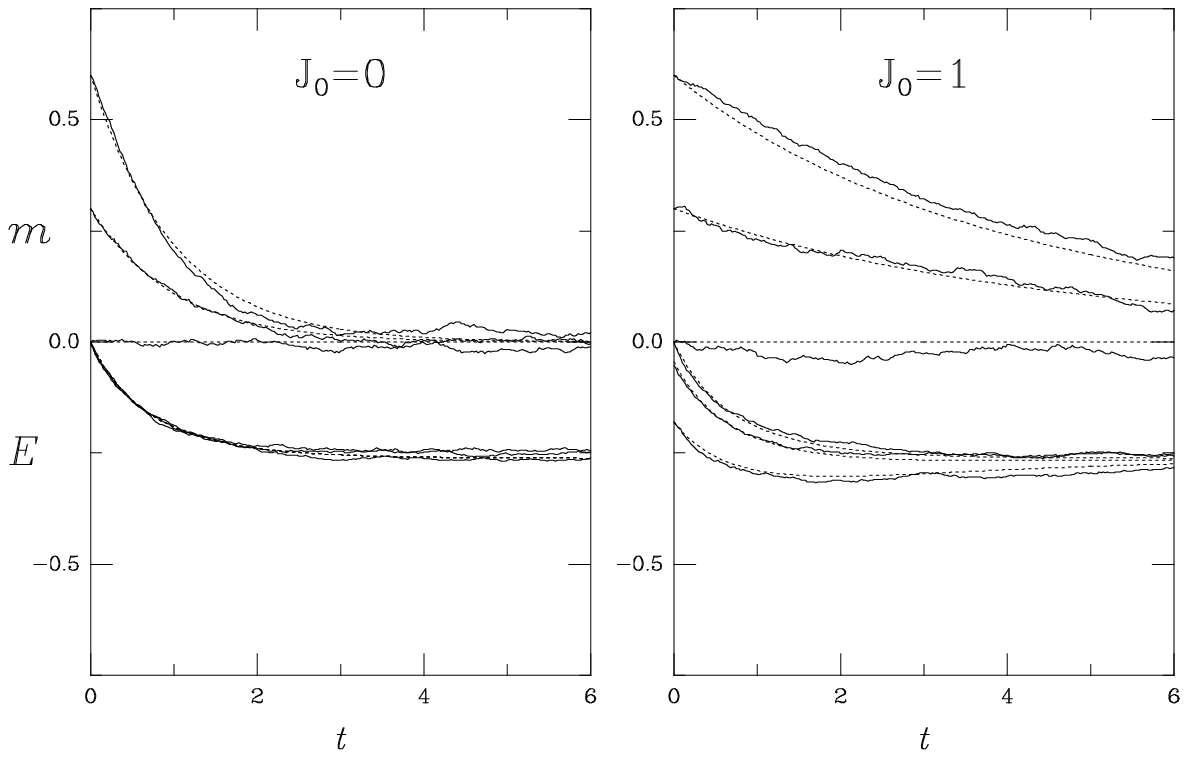}\hspace*{0cm}}
\vspace*{-5mm}
\caption{Magnetisation $m$ and energy
per spin $E$ in the asymmetric SK model at $T=0$, for $J_0=0$ (left) and
$J_0=1$ (right) . Solid lines:
numerical simulations with $N=8000$; dotted lines: result of solving
the RS diffusion equation.}
\label{fig:aflowT0}
\vspace*{105mm}
\hbox to
\hsize{\hspace*{-0cm}\includegraphics{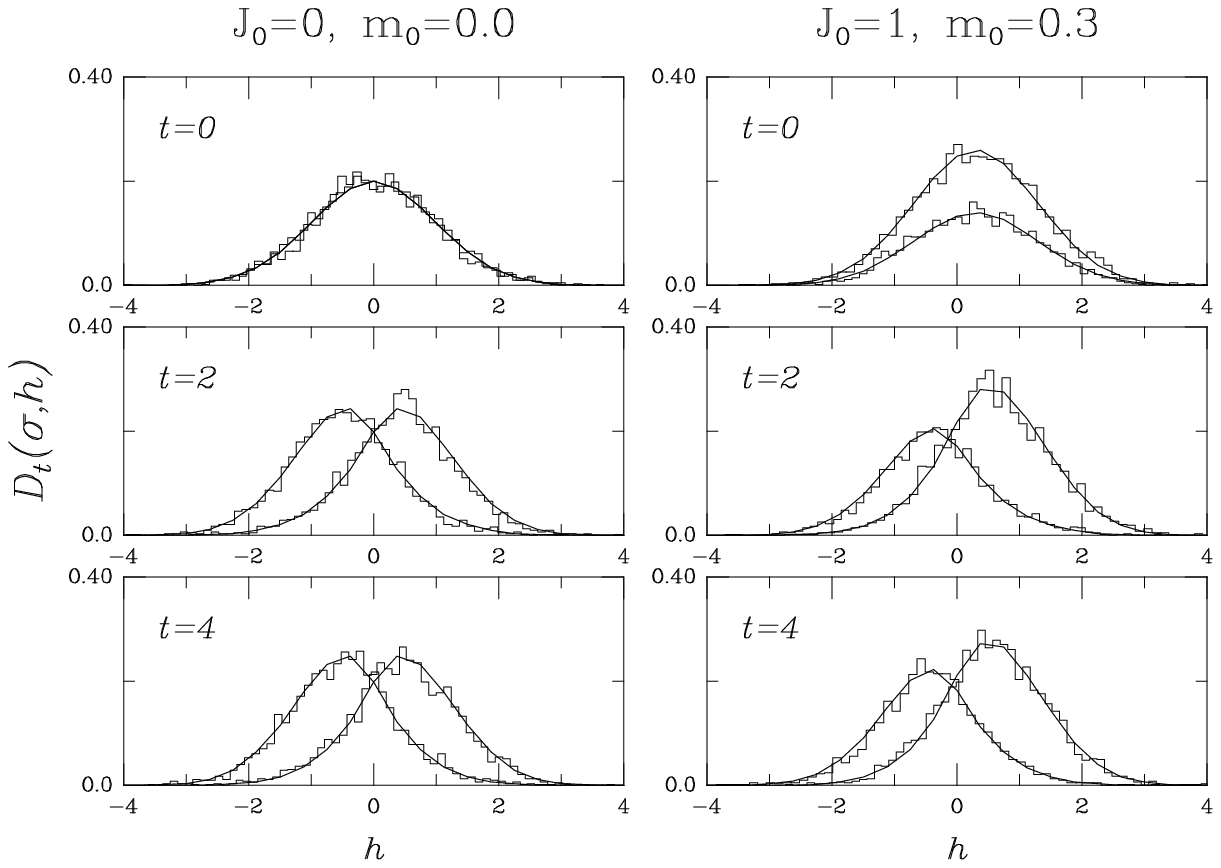}\hspace*{0cm}}
\vspace*{-5mm}
\caption{Field distribitions $D_t(\sigma,h)$ in the asymmetric SK model
at $T=0$, for $J_0=0$ (left) and $J_0=1$ (right) . Histograms:
numerical simulations with $N=8000$; lines: result of solving
the RS diffusion equation.}
\label{fig:adistT0}
\end{figure}
\clearpage
We test our theory by comparing the
results of solving numerically
equation (\ref{eq:finaldiffusion}) in RS ansatz, for the two
model choices $\omega=0$ and $\omega=\frac{1}{2}\pi$, with the results of
performing numerical simulations of the discretised version
of the stochastic dynamics (\ref{eq:master},\ref{eq:ratesandfields})
in a system of size $N=8000$.
Since solving
(\ref{eq:finaldiffusion})
requires a significant computational effort, even within the RS
ansatz, we restrict our experiments to zero external fields and to
initial
configurations  with spin states
chosen independently at random, given a required initial
magnetisation.
For the original SK model,
obtained upon making the choice
$\omega=0$,
the results of confronting
(\ref{eq:finaldiffusion}) with typical simulation
experiments at $T=0$
are shown in figures
\ref{fig:flowT0} and \ref{fig:distT0}, for $J_0=0$ (left pictures) and
$J_0=1$ (right pictures). In figure \ref{fig:flowT0} the top graphs
represent the magnetisation $m$ and the bottom graphs
represent the energy per spin $E$; for the two initial conditions
$m_0=0$ and $m_0=0.3$.
Figure \ref{fig:distT0} shows the corresponding distributions
$D_t(\sigma,h)$ for one particular choice of initial state ($D_t(1,h)$: upper
graph in
$t=0$ window, right graph in $t>0$ windows; $D_t(\minus 1,h)$: lower
graph in $t=0$ window, left graph in $t>0$ windows).
For $J_0=1$ we were not able to calculate the solution of equation
(\ref{eq:finaldiffusion}) up to $t=6$, due to the critical behaviour
of the saddle-point equations.
For the fully asymmetric model, corresponding to
$\omega=\frac{1}{2}\pi$, one finds much simpler equations, due to a
decoupling of the spins from the fields.
In this case equation
(\ref{eq:finaldiffusion}) in fact allows for distributions
$D_t(\s,h)$ which remain of a Gaussian form at all times, in accordance
with  \cite{crisantisompolinsky,riegeretal}.
The results of confronting
(\ref{eq:finaldiffusion}) with $T=0$ simulation
experiments for the asymmetric $\omega=\frac{1}{2}\pi$ SK model
are shown in figures
\ref{fig:aflowT0} and \ref{fig:adistT0}, for $J_0=0$ (left pictures) and
$J_0=1$ (right pictures); for the three initial conditions
$m_0=0$, $m_0=0.3$ and $m=0.6$.
Figure \ref{fig:adistT0} shows the distributions
$D_t(\sigma,h)$ for one particular initial state ($D_t(1,h)$: upper graph in
$t=0$ window, right graph in $t>0$ windows; $D_t(\minus 1,h)$: lower
graph in $t=0$ window, left graph in $t>0$ windows).
\vsp

In this letter be have shown how one can solve the dynamics of
disordered spin systems on finite time-scales with a dynamical
generalisation  of equilibrium replica theory.
Although we have restricted our analysis by making
the replica-symmetric (RS) ansatz,
on the time-scales considered
the agreement between theory and simulation experiment is already quite
satisfactory.
At this stage we  need more efficient numerical procedures in order
to extend the time-scales for which we can solve the equations of the
theory. This would enable us to compare, for instance, with data such
as the ones in \cite{kinzel}, and to investigate the possible existence of
stationary states other than the one corresponding to thermal
equilibrium.
Our theory is by construction exact in various limits. Its full
exactness depends crucually on whether the joint spin-field
distribution indeed obeys a closed self-averaging dynamic equation, which is
difficult to verify.
We plan to investigate several approaches
to this problem in the near
future. Firstly, we want to apply our formalism to
disordered spin systems for which the dynamics has been solved by
other means, like the toy model
\cite{CF}, or the spherical spin-glass
\cite{cugliandolodean}.
Secondly we want to try to derive a diffusion equation for the joint
spin-field distribution, starting from the exact equations for
correlation- and responsefunctions, as obtained from the path-integral
formalism. The latter approach involves (rather
complicated) closed equations
for the correlation function $C(t,t^\prime)$ and the response function
$R(t,t^\prime)$, with two
real-valued arguments each (two times). The present
formalism also involves two functions $D_t(1,h)$ and $D_t(\minus 1,h)$, with
two real-valued arguments each (one time and one field).
It is therefore quite imaginable that both formalisms
constitute exact discriptions of the dynamics of disordered spin models.

\end{document}